\documentclass{elsarticle}
\usepackage{graphicx} 
\begin{document}
\title{The Revolution Has Arrived: What the Current State of Large Language Models in Education Implies for the Future}
\author{Russell Beale}
\date{March 2025}
\begin{abstract}
Large language Models have only been widely available since 2022 and yet in less than three years have had a significant impact on approaches to education and educational technology.  Here we review the domains in which they have been used, and discuss a variety of use cases, their successes and failures.  We then progress to discussing how this is changing the dynamic for  learners and educators, consider the main design challenges facing LLMs if they are to become truly helpful and effective as educational systems,  and reflect on the learning paradigms they support.  We make clear that the new interaction paradigms they bring are significant and argue that this approach will become so ubiquitous it will become the default way in which we interact with technologies, and revolutionise what people expect from computer systems in general.  This leads us to present some specific and significant considerations for the design of educational technology in the future that are likely to be needed to ensure acceptance by the changing expectations of learners and users.
\end{abstract}

\maketitle

\section{Introduction}
Large Language Models (LLMs) – massive deep learning neural networks trained on vast text corpora – have rapidly emerged as powerful tools for education: the public release of models like OpenAI’s ChatGPT in late 2022 catalyzed global interest in applying LLMs in classrooms and curricula. Educators and researchers are exploring how these AI systems can revolutionize teaching and learning across primary schools, secondary education, and universities. Early commentaries heralded LLMs’ potential to generate content, engage learners in dialogue, and personalize learning at scale, while also warning of challenges around accuracy, bias, and academic integrity (Kasneci et al., 2023)\cite{Kasneci2023}. Since then, a fast-growing body of research – spanning computer science, education, and learning sciences – has begun to systematically investigate LLM-driven educational innovations. Recent surveys provide overviews of this nascent field\cite{Xu2024Survey}\cite{Wang2024Survey}, and meta-analyses of experiments are even emerging to gauge impacts on learning outcomes (Deng et al., 2024)\cite{Deng2024Meta}. Building on these works, this paper synthesizes current global research on LLMs in education, focusing on new opportunities they offer, novel techniques and pedagogies they enable, and the implications for learners, teachers, and educational institutions. We highlight key research findings, discuss how the field has evolved over the last few years, and identify emerging trends. We also examine sociotechnical perspectives – such as equity, learner perceptions, trust, and the changing role of teachers – which are crucial for responsible integration of LLMs into education. We then look at the implications that this shifting dynamic in the classroom and outside will have on learner expectations and attitudes, and evaluate those significant implications for the designers of the next wave of technologies.  Overall, LLMs are poised to influence virtually every aspect of education, from daily classroom interactions to curriculum design, but their bigger impact will be on rdefining what the defalt approach to interactions with technology are expected to be. This report is organized as follows: first, we survey the educational opportunities and applications unlocked by LLMs; next, we consider how LLM-based tools are changing learner engagement and educator roles; we then discuss key design challenges and sociotechnical considerations; finally, we reflect on the upcoming challenges.

\section{New Opportunities and Applications Enabled by LLMs in Education}

LLMs are opening exciting avenues in education by enabling forms of instruction and support that were previously impractical or impossible. Unlike prior generations of educational software (which were often limited to canned responses or domain-specific scripts), modern LLMs can engage in open-ended, contextually rich dialogue with users, and this capability allows them to function as flexible conversational agents that can adapt to a wide range of educational tasks. Below, we discuss several major applications and novel techniques for using LLMs across school and higher education.

\subsection{Intelligent Tutoring and Personalized Learning Support}

One of the most promising uses of LLMs is as AI tutors that provide one-on-one support for learners. Ever since the early days of AI in education, the ideal of personalized tutoring – tailoring instruction to each student’s needs – has been a holy grail, and LLMs now make this far more attainable. Models like GPT-4 can converse naturally with students, answer their questions, and adapt explanations based on a student’s progress. Early studies report that generative models can indeed deliver effective tutoring interactions - in one case, a conversational AI built on GPT was used to provide personalized maths instruction; it generated feedback tailored to each student’s errors, leading to improved learning outcomes\cite{cai_bandit_2021}, and mimics the scaffolding provided by human tutors.  With LLMs, such adaptive dialogues can potentially be extended to many subjects and age groups. Students can essentially have a 24/7 virtual tutor that responds in real time, allowing learning to continue beyond the classroom, especially valuable in overcrowded classrooms or under-resourced contexts where teachers cannot give extensive individual attention. Recent work by Kasneci et al. (2023)\cite{Kasneci2023} highlights that LLM-powered agents can “improve student engagement and interaction, and personalize learning experiences” by responding to individual queries and learning pace. Case studies like Khan Academy’s Khanmigo (built on GPT-4) exemplify this, where an LLM tutor guides students through problems using a Socratic approach – asking probing questions rather than simply giving away answers – thereby keeping the learner actively involved. Such AI tutors can also take on personas or roles (e.g. a friendly coach, or a debate opponent in a history lesson) to make learning more interactive and fun. Early classroom pilots report that students often appreciate the on-demand help and the non-judgmental nature of an AI tutor, which can increase their confidence and autonomy in learning. Teachers involved in these pilots observed students becoming more self-directed – for instance, attempting challenging tasks knowing that AI help is available if needed - which suggests LLM tutors, if used appropriately, can foster independent learning strategies. However, realizing the full potential of AI tutors requires careful design. Researchers are experimenting with techniques like prompt engineering and reinforcement learning to ensure the tutor asks good questions and adapts correctly. For example, one technique is to have the LLM follow a “step-by-step reasoning” chain-of-thought and explicitly check the student’s response at each step \cite{parker_large_2024}, which can help the AI provide more coherent guidance. Another approach is using multi-armed bandit algorithms to dynamically adjust the tutor’s feedback strategy based on what seems to help the student most (as in Mui et al.’s work)\cite{Mui2021}, to to optimize the personalization provided by the LLM.

LLMs offer a leap forward in intelligent tutoring systems (ITS) since their open-domain conversational ability , combined with pedagogical strategies, can deliver highly individualized support. This is not limited to factual Q\&A – tutors can engage students in reflective dialogue, ask them to explain their reasoning (learning by teaching the AI), or role-play real-world scenarios. Early evidence is encouraging; a systematic review of experimental studies found that LLM-based interventions “improve academic performance, affective-motivational states, and higher-order thinking propensities” on average (Deng et al., 2024)\cite{Deng2024Meta}. Notably, these benefits were observed largely at the university level (where most studies have been done so far), often in language learning and writing tasks. Going forward, extending such personalized AI tutoring to younger learners and different domains (STEM, humanities, arts) is an active area of research. LLMs can provide one-on-one tutoring at scale, potentially democratizing access to personalized learning. As one education technology leader noted, if we achieve this, "We’d have an Aristotle for an army of Alexanders." \cite{Campbell2019} ----  every student can have personal one-to-one teaching from Aristotle, or Hawking, or Piaget --- an idea that could profoundly change education.  Not only is this an exciting concept, it is educationally significant.  As Bloom noted \cite{bloom1984} back in 1984, his 2-sigma concept refers to the phenomenon that the average student tutored one-to-one using mastery learning techniques (i.e. step-by-step, progressing only when expert at that level) performed two standard deviations better than students educated in a conventional manner. i.e. personal tutoring provides an average 98\% above the level of their colleagues, and 90\% of the students tutored this way achieved results that only 20\% of conventionally taught students could achieve.  Put simply, one-to-one tuition can make everyone a prodigy. That goal is within reach.

\subsection{Content Generation and Instructional Design}

LLMs are also transforming how educational content is created and used, offering powerful assistance to teachers, instructional designers, and even students. Because LLMs can generate human-like text on virtually any topic, they serve as content creation tools that can save educators considerable time and enable new pedagogical approaches. For instance, teachers can prompt an LLM to generate quiz questions, example problems, or even full lesson plans tailored to specific learning objectives. Routine tasks like composing practice exercises of varying difficulty or writing explanation prompts can be offloaded to an AI assistant, which allows teachers to focus more on higher-level instructional design and on interacting with students, rather than spending hours crafting materials. In a survey of early adopting educators, many reported using ChatGPT to create assessment questions and worksheets aligned with their curriculum, finding that it improved efficiency and sometimes the quality of questions by providing fresh variations (e.g. rephrasing problems in new contexts)\cite{baidoo-anu_education_2023}\cite{zhai_chatgpt_2023}. LLMs can rapidly generate multiple-choice questions, open-ended discussion prompts, or case studies, which teachers can then review and curate. Beyond assessments, LLMs can help generate lesson outlines, explanations, and examples. For example, a science teacher could ask the model to produce an age-appropriate explanation of a complex concept (like photosynthesis for 5th graders) or to suggest an analogy to use in class. The model might respond with a creative analogy (e.g. comparing a cell to a factory) that the teacher can refine and present. This kind of AI brainstorming can inspire teachers and enrich the repertoire of teaching strategies. Some educators have also used LLMs to draft lesson plans or lecture notes, which they then customize – effectively using the AI as a first-draft generator or an “idea partner”. Research by Rudolph et al. \cite{rudolph_chatgpt_2023} suggests that teachers leveraging ChatGPT for planning were able to adopt flipped classroom models more easily, by quickly generating reading guides and interactive in-class activity prompts. The AI-generated materials freed up time for the teacher to devise how to use class time for deeper engagement, thus encouraging a shift in pedagogy \cite{radeva_guest_2024}.

Students, too, can benefit from content generation capabilities. For instance, an LLM can simplify or translate content to make it more accessible (useful for language learners or students with learning difficulties). It can generate summaries of complex readings, create flashcards for review, or even produce reading comprehension questions on the fly. These uses align with strategies for active studying – a student can ask the AI to quiz them or to explain a difficult paragraph in simpler terms, facilitating self-study. Another novel technique is using LLMs to generate adaptive texts or scenarios for project-based learning. Suppose students are doing a role-play exercise in a history class; the teacher could use an LLM to generate a realistic historical scenario or dialogue that students then act out or respond to.  For example, we have created a William Shakespeare chatbot that answers questions about his plays, his life and attitudes, in a playful manner to drive engagement with Shakespeare in local schoolchildren. In language learning, teachers have used LLMs to create dialogues or stories in the target language at an appropriate proficiency level. There are already specialized tools incorporating LLMs for such purposes. For example, Wang et al. \cite{wang2024largelanguagemodelseducation} discuss two models: Curipod uses an LLM to generate entire slide decks with interactive elements (polls, short answer questions, etc.) on a given a topic, which allows instructors to produce engaging lecture materials with minimal effort, whilst another, QuestionWell, generates a large bank of questions on any topic with a single prompt, serving as a starting point for quiz or exam creation. Early adopter feedback on these tools is mixed – while they greatly speed up content creation, educators must carefully vet the AI output for correctness and alignment with learning goals. Nonetheless, as the technology matures, we can envision AI-assisted curriculum design becoming routine. A teacher might input the week’s objectives and readings, and the AI suggests slide content, examples, and formative questions for each lesson. Such collaboration could lower the preparation workload on teachers and also provide them with new ideas (reducing “blank page” syndrome). 

From a pedagogical perspective, AI-generated content can support differentiated instruction. Because an LLM can adjust the difficulty or style of text, teachers could generate multiple versions of an explanation to cater to students at different reading levels. Similarly, an AI could produce extension problems for advanced learners and simplified practice for those struggling, all based on the same core material. This addresses the perennial challenge of mixed-ability classrooms by offering tailored resources for each student group. However, reliance on AI content generation requires vigilance. Design guidelines emphasize that teachers should remain the final curators of content to ensure accuracy and appropriateness. There have been cases of LLMs generating incorrect or biased information confidently – if a teacher were to distribute such content unchecked, it could mislead students. Researchers are exploring techniques to mitigate this: for example, prompting the LLM to cite sources for factual claims, or using knowledge-grounded LLMs that refer to a trusted textbook or database when generating educational text. There is also interest in using LLMs in combination with knowledge graphs or domain models to keep the content aligned with canonical knowledge (thus avoiding hallucinations). Despite these caveats, content creation is clearly a domain where LLMs can offer immediate practical benefit. A recent systematic review by Labadze et al. (2023)\cite{Labadze2023ChatbotsSLR} found that “for educators, the main advantages [of AI chatbots] are time-saving assistance and improved pedagogy”, echoing the anecdotal reports that AI helps teachers do more in less time. Thus LLMs are becoming versatile assistants for instructional design – analogous to an ever-ready teaching aide that can produce an initial draft of almost any material, and when guided by skilled educators, these models can enhance the diversity and adaptability of learning resources available to students.

\subsection{Assessment and Feedback}

Assessment is another critical area in education that LLMs are beginning to reshape. Traditionally, providing timely and personalized feedback to students – on essays, open-ended responses, coding assignments, etc. – is extremely resource-intensive. LLMs offer the ability to automatically evaluate and give feedback on student work in ways that go beyond earlier grading software. For instance, LLMs can read a student’s essay and provide comments on coherence, argument strength, and writing style, much like a human reviewer. Studies have found that GPT-based models can grade short essays with reasonably high agreement to human graders and even provide feedback comments that students find useful (e.g. pointing out unclear sentences or suggesting additional evidence), although some depth is lacking compared to expert instructors\cite{alrashidi2022evaluating}. As a result, some institutions are piloting LLMs as a support tool for teachers in grading –-- the AI generates an initial assessment, which the teacher then reviews and adjusts. This can significantly cut down grading time while still ensuring a human in the loop for quality control \cite{Lu2021}\cite{liang_automated_2018}.

Beyond grading, the feedback generation capabilities of LLMs enable new forms of formative assessment. For example, a student writing a report can periodically ask an LLM to critique their draft or identify gaps in their argument. The LLM might respond with something like: “Your introduction is engaging, but I’m not sure the thesis is clearly stated. Also, have you considered addressing counterarguments?” – feedback that can help the student improve before the final submission. This immediate, iterative feedback loop was difficult to achieve at scale before. Preliminary research indicates that students who used an AI writing assistant for feedback made more revisions and produced slightly higher-quality writing on average, though results vary \cite{Marzuki2023}. Another emerging application is automated question-answering and homework help. Students can ask an LLM to check their work or to provide hints on homework problems. Rather than giving the final answer (which would defeat the purpose of the assignment), a well-designed system can deliver Socratic hints. For instance, if a student solved a physics problem incorrectly, the LLM might analyze the work and point out, “I notice you assumed air resistance was negligible – what if that assumption is removed?” or guide the student to revisit a certain equation. This effectively provides personalized formative assessment for each step of a student’s solution.  

One study on high-school maths problem solving showed that with proper prompting, a generative model could analyze student solutions and generate hints that helped students arrive at the correct answer on their own \cite{Barana2023}. The benefit is twofold: students get immediate feedback when practising at home (instead of waiting days for the teacher to mark their work), and teachers get insights into common errors if the AI reports analytics on where students needed hints. LLMs are also being used to generate test and quiz content, which overlaps with instructional design but specifically aids assessment. They can produce large pools of practice questions, including open-response questions that test deeper understanding. Some tools use LLMs to generate adaptive quizzes – the next question’s difficulty is adjusted based on the student’s performance on previous ones, aiming to find the edge of the student’s understanding. This adaptability, powered by an on-the-fly analysis by the model, is a novel technique for personalized assessment that traditionally required complex algorithms and ample item data. In higher education, particularly large courses, LLMs have been trialled to handle routine Q\&A and even grading of discussion posts. For example, an instructor of an online course might use an LLM to monitor the course forum: if a student’s question has been answered many times before, the AI can instantly respond with a synthesized answer and links to relevant resources, acting as a first-line teaching assistant. If the AI is unsure or the question is novel, it flags for a human instructor. This approach was explored by Sajja et al.  \cite{sajja_artificial_2024} who found that an “AI TA” could resolve many student queries satisfactorily, reducing the load on educators and providing faster responses to students. 

Perhaps one of the most innovative assessment applications is using LLMs to enable open-ended, authentic assessments that were previously hard to grade. For instance, instead of a traditional exam, a professor might assign students to have a structured debate with ChatGPT on a topic, then evaluate the transcript quality. The rationale is that if the AI can simulate a debate opponent or a collaborative partner, the interaction itself becomes an assessment of the student’s skills (how well they reason, use evidence, correct the AI when it’s wrong, etc.). LLMs can play multiple roles in such scenarios – both generating the prompts and participating in the dialogue. While still experimental, this points to a future where assessments could be more dialogic- and process-oriented, rather than just static outputs, with LLMs helping to facilitate and evaluate the process.

It should be noted that this initial evidence regarding LLM-driven assessment is mixed. On the positive side, meta-analyses suggest that students using AI feedback tend to show higher engagement and sometimes better performance\cite{Deng2024Meta}. On the cautionary side, concerns about over-reliance are significant: if students lean too much on AI for homework, they might bypass important learning (discussed  in Section \ref{sec:designchallenges}). Likewise, if teachers trust AI grading without verification, errors or biases in feedback could adversely impact students. Researchers like Li et al. (2024)\cite{li_ethical_2024} identify academic integrity as a “predominant focus” in current ChatGPT-in-education research – a reminder that assessment applications must be designed very carefully to avoid facilitating cheating or giving unfair advantages. One promising safeguard is to use LLMs to augment peer assessment rather than replace instructor assessment. For example, students can use an LLM to get preliminary feedback on each other’s work, which can improve the quality of peer reviews - if a student is unsure how to critique an essay, the AI can suggest some points. This maintains human involvement while using AI to scaffold the process.

LLMs are enabling more immediate, personalized, and diverse forms of feedback and assessment. From automating parts of grading, to giving students on-demand hints, to creating new interactive assessment formats, these tools have the potential to both enhance learning and streamline teaching. The key is to integrate them in a pedagogically sound way – using AI to support learning without undermining the development of student skills and without compromising fairness. When done right, LLMs can help shift assessment towards a more formative, learning-oriented approach rather than solely summative judgments.

\subsection{Collaborative and Creative Learning with AI}

Beyond one-on-one tutoring and content delivery, LLMs also open up possibilities for collaborative and inquiry-based learning experiences. Rather than viewing the AI as just a teacher or evaluator, we can also cast it as a participant in learning activities – a kind of intellectually capable peer that can stimulate discussion, creativity, and exploration. This is leading to novel pedagogical approaches where students learn with the AI as a partner. One example is using an LLM as a conversational partner for practising communication skills. In language learning classrooms, students have started conversing with chatbots to practice foreign languages. The LLM can play the role of a conversation partner who never gets tired or judgmental. For instance, a student learning English can chat with the AI about their day, and the AI can correct grammar mistakes or teach new vocabulary in context. This provides immersion and practice opportunities beyond what the teacher alone could provide, especially in large classes. Similarly, for developing argumentation or debate skills, students can engage in debates with ChatGPT on various topics. Because the AI can intentionally take a counterpoint, even a lone student can practice debating both sides of an issue. Early studies on AI-driven dialogue practice show promise: one experiment had students debate ethical dilemmas with an AI, and the students reported it forced them to think more deeply and consider perspectives they hadn’t before \cite{Lee2023debate}. LLMs can also facilitate collaborative learning among students by acting as a mediator or coach in group discussions. Consider a scenario where a small group of students is working on a problem. The AI can be present in the group chat to gently guide the discussion if it stalls – e.g., by posing a question (“Have you considered the effect of X on your solution?”) or suggesting a conflict resolution if students disagree. Essentially, the AI is like an ever-available moderator that can keep collaboration productive. Research in computer-supported collaborative learning (CSCL) is exploring this, with some prototypes showing that groups with an “AI facilitator” tended to stay more on task and cover problem facets more thoroughly than control groups without it\cite{sahab2024conversational}.

Another collaborative dynamic is students and AI co-creating artifacts together. For instance, in a creative writing class, a student could write a short story by alternating turns with an AI: the student writes one paragraph, the AI writes the next, and so on. This kind of human-AI co-writing can spark student creativity (the AI might introduce an imaginative twist the student hadn’t considered) and also compel the student to critically evaluate the AI’s contributions (learning to discern which ideas to accept or reject)\citeyear{woo_writing_2023}. A study by Zhao \citeyear{zhao_impact_2024} found that high-school students who co-wrote stories with an AI produced more innovative and richer narratives, and recognised its benefits as a form of distributed cognition supporting their creative practice. Similarly, in coding education, a student can collaborate with an AI coding assistant (like GitHub Copilot) to develop a program, learning new coding techniques from the AI’s suggestions while practicing debugging and problem decomposition. This “pair programming with AI” \cite{ma_is_2023} approach has been reported to improve novice programmers’ confidence, though care is needed to ensure they don’t rely on the AI for every solution.

Project-based learning can also leverage LLMs. Imagine a history project where students must interview historical figures – an LLM can simulate a figure (say, Marie Curie or Nelson Mandela) and students conduct a mock interview by asking questions and getting answers in character. This engages students in inquiry: they must formulate good questions and also verify the AI’s answers against factual sources (since the AI might not be perfectly accurate). It blends creativity with research, as students critique the AI’s “historical” responses\cite{schoolai2025}\cite{mintz2025}. Through such activities, LLMs can act as a multipurpose role-player – be it a historical person, a virtual lab partner, or a character in a scenario. This enables simulations and role-plays that enrich learning in social studies, science, and literature classes. Students can conduct science experiments in conversation with an AI playing the role of a lab assistant, or re-enact a scene from a Shakespeare play with the AI taking one of the roles, etc. These imaginative uses can increase student motivation and engagement, as they introduce an element of novelty and play into learning. Collaborative learning with AI also raises interesting dynamics about how students perceive the AI as a social entity.

Some research is examining learner perception of AI partners – do students treat the AI more like a tool, or do they attribute some kind of peer-like status? Jeon \& Lee \citeyear{Jeon2023TeacherChatGPT} investigated teacher–AI collaboration and identified roles, which can parallel student experiences: in their study, ChatGPT took on roles such as an “interlocutor” (conversational partner) and “teaching assistant,” and teachers valued it as a collaborator but still recognized its non-human nature. With students, early observations show that many enjoy interacting with the AI but also personify it to some degree (e.g., thanking it, or feeling frustrated at it when it “doesn’t understand”). This can actually be used pedagogically – for instance, prompting students to critically reflect: “Do you think the AI understood you? Why might it have given that response?” – leading into discussions on how the technology works (i.e., developing AI literacy alongside subject matter learning). From a sociocultural perspective, incorporating an AI into group work can shift the classroom norms, sometimes for the better: some educators note that students who are shy to speak up in class may find it easier to first express their ideas to the AI tutor/agent, and then bring those ideas to the human group, having gained confidence. Conversely, there’s a risk that more assertive students might dominate interactions with the AI, or that the AI’s suggestions could override quieter students’ ideas. Designing equitable collaboration setups will be important (perhaps by rotating which student interacts with the AI, etc.).

We are only beginning to understand the best practices for AI-supported collaborative learning. Nonetheless, it is clear that LLMs can play many roles: coach, teammate, devil’s advocate, simulator, etc. These roles, when woven into pedagogy thoughtfully, can create rich learning experiences. They align well with constructivist and constructionist approaches – students actively constructing knowledge through dialogue and creation, with the AI expanding what’s possible. For example, students can pursue inquiries that go beyond the textbook by asking the AI to explain advanced topics or connect ideas, thus pushing the boundaries of the curriculum based on their curiosity. 

LLMs are not just tutors or content generators; they can be participants in the learning process. This opens the door to more interactive, collaborative, and creative learning strategies. Classrooms of the future might regularly see human students and AI agents working side by side on problems, learning from each other. The teacher’s role in such settings becomes one of orchestrating these interactions to ensure they are productive and inclusive. We now turn to exactly that point – how LLMs are influencing the roles of teachers and learners, and what new models of instruction are emerging.

\section{Changing Learner Interaction and Educator Roles}

The introduction of LLMs into educational settings is not only creating new opportunities, but also shifting the dynamics of classrooms and learning. In particular, the ways learners interact with educational content (and with instructors) are evolving, and the role of human teachers is being reimagined in the presence of AI assistants. In this section, we examine these transformations. We first look at how LLM-based tools influence student engagement and behaviour, then consider the perspectives of teachers and how pedagogy and teacher roles are adapting.

\subsection{Learner Interaction, Engagement, and Perceptions}

Students today are a digitally native generation, and many have eagerly embraced AI tools like ChatGPT for learning – sometimes to the surprise or concern of their instructors. Global surveys conducted in early 2023 found a wide range of student reactions: some viewed ChatGPT as a “fantastic study aid” that can clarify concepts or help with writing, while others were wary about its accuracy or worried that using it might be considered cheating \cite{Ansari2023ChatGPTScoping}. On the whole, students tend to appreciate the immediacy of LLM support which leads to increased engagement in independent study. Rather than hitting a roadblock and giving up or waiting to ask the teacher the next day, students with access to an AI tutor often persist longer on tasks, consulting the AI for hints. This on-demand assistance has been linked to  improved attitudes towards  assignments in some studies \cite{fuchs2023}. Furthermore, learner–AI interaction is typically in natural language (typing or speaking questions and getting explanations). This conversational mode can make learning feel more interactive and “alive” compared to reading a static textbook. Students have reported that “learning by chatting” with an AI feels less formal and more personalized – they can ask very specific questions without fear of judgment. Over time, as students regularly use an LLM, they may even develop a kind of mental model of the AI’s capabilities and quirks. For example, a student might learn that “the AI sometimes gives too much info, so I need to ask more directly” or “I should double-check maths answers it gives.” In other words, through experience, students refine their strategies for querying the AI (developing their AI and prompt literacy skills). This metacognitive aspect is important: ideally, interacting with AI can help students become more reflective about their own understanding (“What exactly do I need to ask to overcome my confusion?”).

On the engagement front, initial research indicates that LLMs can both increase cognitive engagement and pose risks of surface engagement. On one hand, tools like AI tutors that ask follow-up questions push students to explain their reasoning, thereby engaging them in deeper thinking. On the other, if a student uses ChatGPT to do their homework by simply copying its answers, then engagement is superficial and learning can suffer. This dichotomy is evident in usage patterns observed: many students start by using AI as a crutch (just getting answers), but with proper guidance and honour codes, they can be encouraged to use it as a learning tool (getting hints, verifying answers, exploring alternate solutions). Some educators have implemented structured activities to channel student–AI interaction productively. For instance, a professor might allow students to consult ChatGPT during an assignment but require them to submit a short reflection on what they asked, how they used the answer, and what they learned. Such practices nudge students to engage more meaningfully and honestly with the AI. In terms of learner perception and trust, students are learning to calibrate their trust in LLM outputs. Early on, many students assumed the AI is always correct (it sounds confident and authoritative). This led to instances of learners internalizing false information or incorrect problem-solving methods. However, as the novelty wears off and as instructors raise awareness of AI limitations, students are becoming more sceptical and evaluative of AI responses – a positive development. In fact, using LLMs can teach critical thinking, because students are often quick to discover that “not everything ChatGPT says is true.” When a student encounters a mistake from the AI, it presents a learning opportunity: why was it wrong? How can I verify answers? Some teachers deliberately give assignments where one task is to find and correct the AI’s errors. For example, providing an AI-written essay that contains some factual inaccuracies or logical fallacies, and asking students to critique it. Students generally find this exercise engaging (perhaps more so than critiquing a peer’s essay) and it trains their critical reading skills. This risk of over-trust among some learners is especially true for younger ones who might not have the skills to assess accuracy. If an AI explanation sounds good, they might accept it uncritically. Addressing this, Kasneci et al. (2023)\cite{Kasneci2023} argue that education systems must now include fostering AI literacy: teaching students how these models work at a basic level, their tendency to hallucinate, and strategies to cross-check information. Indeed, building trust with discernment is key: we want students to trust AI enough to use it as a helpful resource, but not so blindly that they believe everything it says.

Learner engagement is also influenced by the novelty and interactiveness of LLM tools. Many students initially find using an AI “cool” or motivating. Usage statistics from pilot programs show high voluntary usage in the first few weeks. There is concern, however, about the novelty effect wearing off. Will students still be as engaged when the AI becomes a normal everyday tool? Or might they get lazy and let the AI do more of the work? Deng et al. (2024)\cite{Deng2024Meta} note this explicitly and call for studying long-term impacts, as their meta-analysis found positive motivational effects in the short term but it is unclear if these persist. Some anecdotal evidence suggests that after the novelty fades, students settle into patterns – some using the AI diligently as a support, others perhaps using it only last-minute for answers. Effective pedagogy and clear guidelines can help maintain constructive engagement. For example, educators recommend setting explicit expectations like “It’s fine to use AI for ideas, but you must write the essay yourself,” or “If you used AI to check your work, note that in your submission.” Such transparency removes the stigma and lets students focus on learning gains rather than hiding AI use. A fascinating sociological aspect is how LLMs might change peer interactions. If a student can get help from AI, will they ask their classmates less? Some educators worry about reduced human peer learning. However, early classroom observations show that students often share interesting things they tried with the AI with each other (“Look what ChatGPT suggested for problem 3!”). This can actually spark peer discussions about the AI’s answers, leading to collaborative evaluation of AI outputs. In some cases, students almost treat the AI as a common resource and collectively figure out how to get the best out of it. Thus learner interactions in the age of LLMs are marked by increased agency (students can explore knowledge freely with AI), but also require new skills of judgment and self-regulation. 

When appropriately guided, LLMs can make learning more engaging, interactive, and tailored to individual curiosities – students can ask endless “why” questions that a teacher might not have time for. But the ease of getting answers means educators must emphasize process over answers, and cultivate an inquiring mindset. The student’s role may gradually shift from information receiver to information navigator, with the AI as a compass. Next, we consider the teacher’s role in this evolving landscape, as they are the linchpin in ensuring AI is used effectively and ethically in the classroom.

\subsection{Educator Roles and Instructional Models with LLMs}

The advent of LLMs in education is prompting educators to rethink their roles and pedagogical models. Rather than rendering teachers obsolete (a fear sometimes voiced in media), evidence so far suggests that teachers remain as crucial as ever – but their role may shift more towards designing learning experiences, curating AI use, and providing the human touch that AI cannot. Jeon \& Lee (2023)\cite{Jeon2023TeacherChatGPT} conducted an insightful study where human teachers integrated ChatGPT into their instruction for two weeks and then reflected on the experience. They identified four distinct roles that the AI played (from the teachers’ perspective): \textit{“interlocutor, content provider, teaching assistant, and evaluator.”} Correspondingly, teachers found their own roles shifting to complement the AI: they acted as orchestrators, making pedagogical decisions about when and how the AI is used; as facilitators, encouraging students to be active investigators (e.g., prompting students to fact-check the AI or ask it deeper questions); and as mentors, raising awareness about AI’s ethical use and limitations. This complementary relationship underscores that teachers become more of a guide on the side when AI is in the mix – but an extremely important one who ensures the AI’s contributions actually benefit learning.  In practical terms, teachers are developing new skills akin to AI classroom management. For example, a teacher might need to monitor not just student discussions, but also student–AI interactions to ensure they are on track. If students are chatting with an AI and it goes off-topic or gives a confusing answer, the teacher may need to intervene to clarify or bring the discussion back. Some teachers now prepare prompt templates or examples for their students – effectively teaching them how to ask the AI good questions: a novel instructional task that didn’t previously exist.  Additionally, teachers must set boundaries, deciding which tasks should or should not involve AI. For instance, a teacher might say “During this brainstorming phase you can use ChatGPT for ideas, but in the next phase of writing, turn it off to develop your own thoughts.” Designing these structures for AI usage is becoming part of lesson planning.

A  review by Ansari et al. \cite{Ansari2024} mapped global evidence on ChatGPT’s adoption in higher education. Their work reveals that while many university studies report increased engagement and efficiency in content generation, they also flag important concerns. For example, many instructors in large online courses have experimented with using an “AI teaching assistant” to field routine queries\cite{onyalo_ai_2022, shah_students_2024}.  Other studies have focused on the pedagogical integration and ethical dimensions of LLM use in higher education. Hsiao et al. \cite{hsiao_developing_2023} stresses the need for redesigned assessment frameworks that account for both the capabilities and the limitations of LLMs. In parallel, Guizani et al. \cite{Guizani2025} reviewed practical issues in implementing LLMs in higher education, emphasizing ethical integration, privacy, and the need for interdisciplinary collaboration between AI developers and educators.  Moreover, qualitative studies—such as the work by Jeon and Lee \cite{JeonLee2023}—examine how the role of the teacher adapts when LLMs are used as complementary tools. Their findings indicate that when teachers are actively involved in curating AI outputs and guiding student interactions, LLMs can be a powerful aid. They also highlight concerns: if not carefully managed, AI assistance might lead to superficial engagement or even jeopardize academic integrity if students bypass critical learning steps. Finally, a review by Shahzad et al. \cite{Shahzad2025} encapsulates the overall promise of LLMs for personalized and scalable learning in higher education while also cautioning about risks such as potential bias, issues with transparency, and the possibility of widening the digital divide.

Educators are also playing a key role in establishing trust and buy-in for using LLMs. Many teachers' initial reaction to ChatGPT was concern about cheating. However, as they learn more, some have shifted to an approach of embracing and channelling the technology. They communicate to students that “We know you have access to this tool; let’s talk about what it is good and less good at, and how to use it productively and ethically.” This open dialogue builds trust – students don’t have to hide AI use, and teachers don’t have to play “AI police” as much. Instead, teachers focus on integrating it in a way that upholds academic integrity (for example, requiring personal reflections or oral defenses in addition to AI-assisted written work). In terms of instructional models, one noticeable shift is towards a more facilitative and mentoring teaching style. Since the AI can handle some content delivery or Q\&A, teachers can allocate more class time to higher-order discussions, individualized help, and socio-emotional support. In classrooms where AI tutors are active, teachers often roam and address specific needs – for example, helping a student who is struggling even with AI help, or engaging students in meta-cognitive questions (“What did ChatGPT suggest? Why do you think that is a good/bad suggestion?”). In essence, the teacher becomes a meta-cognitive coach, guiding students to think about their thinking (and about the AI’s thinking). Another emerging model is the “flipped AI classroom.” In a traditional flipped classroom, students first encounter material on their own (e.g., via video) and then do deeper work in class with the teacher. Now, with LLMs, students can interact with the new material at home by asking an AI questions about the video or reading. By the time they come to class, they might have already resolved basic misunderstandings with AI’s help. The class can then focus on more complex applications or discussions. The teacher in this model coordinates with the AI by perhaps providing the AI with the course materials so it can give more context-aware answers to students at home. This is relatively uncharted territory but could be powerful: the AI essentially augments the at-home learning phase, making flipping even more effective.

Teachers are finding a role as quality controllers and editors of AI contributions. For example, an AI might generate a draft lesson or a set of quiz questions – the teacher then curates these, picks the best ones, fixes any issues, and adds the personal context or emphasis needed. This is similar to how teachers curate content from textbooks or the internet, but now the source is AI-generated. It requires strong content knowledge and pedagogical judgement to ensure the AI’s output is used appropriately. This underscores that teacher expertise is still central – the AI can produce content, but deciding what to use and how to use it with students is squarely in the teacher’s domain. Teachers’ workload could shift in nature: less time creating basic materials from scratch, more time designing learning experiences and interventions around AI usage. Importantly, the presence of AI in the classroom allows teachers to dedicate more attention to mentoring and socio-emotional support. A human teacher can build rapport, motivate, and empathize in ways an AI cannot. If some routine tasks are offloaded to AI (like answering “when is the assignment due” or other frequent queries, or giving immediate feedback on a first essay draft), teachers have slightly more bandwidth to check in emotionally with students, to inspire them, to address individual anxieties or interests. In interviews, teachers have expressed that using AI has “freed up more time to interact one-on-one with students who need help” – which is a positive outcome, aligning with the goal that technology should amplify the human element, not diminish it.

Professional development and teacher training programs are beginning to incorporate AI literacy to prepare educators for these new roles. The need for teacher training on LLMs is widely recognized\cite{Kasneci2023}. Educators not only have to learn how to use the tools, but also to understand their limitations (so they can guide students accordingly) and to redesign assessment in ways that account for AI. For instance, some university instructors have redesigned their exams to be more oral or project-based to ensure students can’t just use ChatGPT to cheat – but also to leverage ChatGPT in assignments where appropriate (like using it for initial research but then requiring an in-person presentation). These pedagogical shifts are still experimental, but they reflect teachers actively adapting. A challenge is that not all teachers are comfortable with AI or see its value immediately. Change can be intimidating, and there are valid worries about things like AI replacing some teaching tasks or the teacher losing authority if “the AI knows everything.” However, studies of teacher perception (e.g., Fuligni (2025)\cite{fuligni2025wouldwantaitutor}) show that once teachers try the technology and see it as a helper rather than a threat, many become enthusiastic about weaving it into their practice\cite{kim_teachers_2022}. The key is involving teachers in the co-design of AI tools and pedagogic policies, so that their expertise and needs shape how LLMs are used in the context.

The educator’s role is evolving from primarily delivering content to orchestrating a human–AI hybrid learning environment. Teachers are becoming designers of learning experiences that include AI as a component, coaches who guide students in how to learn (and how to use AI) effectively, and critical filters who ensure that the AI’s contributions serve educational goals. This is encapsulated well by a vision that “AI will not replace teachers, but teachers who use AI may replace those who don’t” - while perhaps hyperbolic, it emphasizes that embracing these tools can enhance teaching. Ultimately, the teacher provides what AI cannot: mentorship, inspiration, ethical guidance, and adaptive understanding of each student’s unique context. The best outcomes seem to occur when teachers and LLMs work in tandem, each doing what they do best – a synergy that can potentially deliver a richer education than either alone.

\section{Design Challenges and Sociotechnical Considerations}\label{sec:designchallenges}

While LLMs bring many promising opportunities to education, they also introduce a host of challenges and complex considerations. Effective integration of LLMs into learning environments requires navigating technical limitations, ethical dilemmas, and social impacts. In this section, we discuss the major challenges identified in the literature, by practitioners, and ourselves. These include issues of accuracy and reliability, biases and equity, academic integrity, privacy, the need for transparency, and the broader sociotechnical impacts on trust and roles. Addressing these challenges is crucial for harnessing LLMs’ benefits without causing harm or exacerbating inequalities\cite{UNICEF2025}.

\subsection{Accuracy, Reliability, and Hallucinations}

LLMs like GPT-4 are powerful but fallible. A well-known issue is that they sometimes generate incorrect information – a phenomenon often referred to as “hallucination.” In educational use, this is a serious concern: a student might earnestly ask the AI a question and receive a confidently stated but wrong explanation (though this is a phenomena not unknown in humans either). If uncorrected, this could lead to misconceptions. Ensuring the accuracy and reliability of LLM-provided information is thus a key challenge. Studies evaluating LLM responses to academic questions have found mixed results. In some domains (especially where training data was rich), the AI answers can be highly accurate, whereas in others (or when questions require nuanced reasoning), errors are common. For example, an experiment by Apornvirate et al. \citeyear{apornvirat_comparative_2024} found ChatGPT scored around 100\% on a set of pathology image interpretation questions when given context, Bard scored around 85\%, but both dropped significantly if context was absent: Bard consistent hallucinated plausible but incorrect answers. In subjects like history or literature, there have been cases of the AI inventing references or misattributing quotes. This unpredictability means that unverified use of LLM outputs is risky in educational settings. One straightforward mitigation is to always keep a human in the loop – i.e. AI as an assistant, not an autonomous source of truth. Teachers and students must verify AI outputs against trusted sources. However, constantly verifying can become burdensome and undermines the efficiency gains. Researchers are actively exploring solutions: one approach is developing education-specific LLMs or fine-tuned models on verified curricular data, which might reduce hallucinations in that context. Another is incorporating retrieval of facts: for instance, tools that have the LLM search a textbook or database and quote it in the response. These retrieval-augmented LLMs can improve factuality by grounding answers in known sources (though they are not foolproof). Another technique is prompting the LLM to show its reasoning (the “chain-of-thought”) which can sometimes make it easier for a user to spot a wrong step if the reasoning is exposed. Still, even with these, a level of uncertainty remains. Therefore, educators often stress to students the importance of double-checking AI-generated information. This challenge also brings an opportunity: as mentioned earlier, it can be turned into a learning experience by engaging students in the verification process. In terms of system design, interface solutions like highlighting text that the LLM is less confident about, or providing multiple possible answers for the user to compare, are being considered to communicate uncertainty. From a deployment perspective, reliability is especially critical in primary and secondary education, where younger students might lack the skills to discern errors. As a result, some schools have initially banned unsupervised use of ChatGPT precisely because of the fear of misinformation. Over time, as better guardrails develop, a more nuanced approach can be taken. Addressing the reliability challenge is essential for trust: if students or teachers frequently encounter incorrect outputs, they may lose trust in the tool altogether (the “once bitten, twice shy” effect). On the other hand, with careful introduction and by building a culture of verification, users can maintain a healthy scepticism without completely losing trust. In fact, part of AI literacy is knowing that these models “are not guaranteed correct” no matter how fluent they sound. Many have called for LLM developers to improve factual accuracy for education use cases; this is a technical frontier, possibly requiring new architectures or training paradigms to reduce hallucination without sacrificing the generative ability.

\subsection{Bias, Fairness, and Equity}

LLMs inherit biases present in their training data and can also exhibit new forms of bias. In an educational context, biased or insensitive AI responses could harm student learning or make some students feel unwelcome. For example, an LLM might inadvertently use stereotypes in examples (always portraying scientists as male, for instance), or it might perform better on content in English than in other languages, disadvantaging non-English speakers. Ensuring equity in AI-supported education is a major concern highlighted in recent research\cite{Wang2024Survey}. Fairness issues can be technical (the model’s outputs) and access-related. The technical side includes the racial and gender biases in some LLM outputs. Access-related ones are around who the AI “understands” best – for instance, if a student writes in a dialect or with non-native grammar, will the AI still effectively help them? If not, that student could get worse support than a more privileged peer. Developers are attempting to mitigate bias via fine-tuning and content filters, but it is an ongoing battle. Educators need to be vigilant: one suggestion is that all AI-supported content for students should be reviewed with a lens of culturally responsive pedagogy - for instance, if an AI-generated set of maths word problems inadvertently only references certain cultural contexts, a teacher could modify them to be more inclusive. Some projects are also exploring custom LLMs for local contexts and languages – for example, an LLM fine-tuned for use in African contexts that includes local names, contexts, and is more attuned to Swahali language usage\cite{noauthor_jacaranda_nodate}. This localization can help reduce cultural bias and increase student identification with the content.

The other equity aspect is access to the technology itself. LLMs (especially the most advanced ones) require significant computing resources and internet access. Students in under-resourced schools or regions with poor connectivity might not be able to use these tools, widening the digital divide\cite{beale_mobile_2008}. There is a real risk that AI in education could exacerbate inequalities: wealthy schools deploy AI tutors and see improved outcomes, while less privileged ones fall further behind. Addressing this requires proactive policy and investment. Some initiatives aim to provide offline or smaller-scale models that can run on local devices, or to negotiate low-cost licenses for educational institutions. Open-source LLMs are also emerging, which could be hosted by schools without relying on expensive API calls, but even free models require hardware and technical know-how to deploy. Governments and international organizations are beginning to recognize this equity issue – for instance, UNESCO’s guidance on AI in education \cite{UNESCO2021} emphasizes inclusion and calls for ensuring all languages and regions benefit from these advances, not just English-speaking wealthy countries. In current research literature, Ansari et al. (2023)\cite{Ansari2023ChatGPTScoping} note that 77\% of studies on ChatGPT in higher education have been in high-income countries. This skew suggests that the voices and contexts of low-income countries are under-represented in shaping how these tools are used. Going forward, more global collaboration and context-specific studies are needed so that LLM integration does not follow a one-size-fits-all approach that ignores local needs.

To ensure fairness within a classroom, teachers may also need to differentiate AI usage. For example, not all students may benefit equally from interacting with a text-based AI – some with reading difficulties might struggle. Perhaps pairing LLMs with text-to-speech or voice interfaces could help such students. Likewise, teachers might allow certain students to use the AI in ways others do not, as an accommodation (similar to how some get calculators or extra time). These decisions must be made carefully to avoid perceptions of unfairness. Overall, tackling bias and equity is not just a technical task, but a socio-technical one: it involves diversifying training data, improving model tuning, expanding access, and consciously integrating AI in a way that narrows gaps rather than widening them. It also means being transparent about these issues with students – e.g., discussing that “the AI may have some biases and here’s why” can be an educational topic itself, fitting into digital citizenship or social science curriculum.

\subsection{Academic Integrity and Ethical Use}

Perhaps the most discussed challenge in popular discourse has been academic integrity. The ease with which students can get an essay or solution from ChatGPT raises obvious concerns about plagiarism and cheating. Many educators and institutions reacted swiftly – some banning AI-generated content or using detection tools. However, detection of AI text is inherently uncertain as AI can often evade detectors, and false positives can harm students wrongly accused. The field is gradually moving towards emphasizing ethical use policies rather than outright bans. Still, the challenge remains: how to prevent misuse of LLMs in assessments and maintain the credibility of qualifications. One strategy is reassessing what and how we assess it. If traditional take-home essays become problematic, more emphasis could be placed on in-class work, oral exams, or portfolio work that includes process documentation. Some instructors now require students to submit drafts, notes, or logs of their work to ensure they didn’t just prompt ChatGPT at the last minute. Others integrate AI openly: e.g., “You may use AI to help, but you must cite any significant contribution from it, and you will be graded on your critical commentary on that AI contribution.” This turns a potential issue into a learning exercise in integrity and proper attribution.

The ethical use of AI extends beyond cheating. It includes concerns like students becoming over-reliant on AI and not developing their own skills (sometimes dubbed the “calculator effect” but for writing/thinking). If a student always uses AI to do the first pass of anything, are they short-changing their learning? This is an open question. Some cognitive scientists worry about skill atrophy – e.g. writing skills or mental arithmetic skills could weaken. Empirical evidence is still sparse on long-term impacts. Deng et al. (2024)\cite{Deng2024Meta} noted no significant effect on self-efficacy in studies so far, but that doesn’t tell us much about actual skill changes. To combat this, teachers can design activities that sometimes intentionally withhold AI or require manual effort first, then allow AI as a secondary step. For example, a teacher might have students write an essay draft by themselves in class without AI, then for homework use AI to get suggestions to improve it, and finally reflect on the differences. This way, students practice the skill and also learn how AI can help refine it – hopefully the best of both worlds.

Another ethical aspect is transparency and attribution; teaching students to be honest about their use of AI is part of academic integrity. Some argue using AI is similar to getting help from a tutor – which is allowed as long as it’s disclosed. Norms may evolve such that “AI assistance” is listed akin to references. The role of teachers is to set clear expectations: e.g., “You may use AI for grammar and style help, but the ideas and wording should be your own, otherwise cite it.” If a student does misuse AI (e.g., submits entirely AI-written work as their own), institutions need policies for that. It is an arms race: as AI improves, it will get harder to distinguish its content from human-generated work. So fostering an ethical culture where students understand that learning (and not just getting a grade) is the goal becomes paramount - and this may require some change in attitude from some educators too. Some educators leverage students’ own sense of professional responsibility: for example, reminding medical students that “In your future practice, you can’t rely on AI blindly – lives are at stake, so you must learn the material.” This can discourage taking shortcuts in learning. Ultimately, academic integrity in the age of AI will likely rely less on policing and more on redesigning curriculum and assessment for authentic learning. If assignments are more individualized, process-based, and emphasize creativity or personal reflection, they become harder to outsource to AI. Moreover, embedding AI literacy and ethics discussions into the curriculum can raise student awareness of why misusing AI is problematic. Interestingly, Kasneci et al. (2023)\cite{Kasneci2023} suggest that bringing AI into education early could help students develop critical awareness of AI-generated content, making them less likely to use it naively or dishonestly. In a sense, if AI is integrated openly with guidance, students might be less tempted to misuse it covertly. They also note that these challenges “are not unique to AI in education” –  plagiarism and misuse have always existed; AI just amplifies the need for robust academic integrity frameworks.

On the educator side, a challenge is how to ethically use LLMs with student data. For example, if a teacher wants to use ChatGPT to help grade essays, they should consider privacy and also whether that is fair to students (are all essays getting equal treatment; what if the AI mis-grades one?). There’s an ethical dimension in the teacher’s use of AI too. Transparency can help here as well – teachers could inform students if AI was used in evaluating their work and allow appeals if something seems off. All these considerations point to the need for clear guidelines and policies at institutional and classroom levels. Many universities have issued interim policies on AI use\cite{noauthor_russell_2023}; professional organizations are formulating codes of conduct. As we collectively learn what works and what pitfalls exist, these guidelines will evolve. The bottom line is that maintaining academic integrity will require a combination of culture-building, pedagogy adjustment, and perhaps technological tools (though playing cat-and-mouse with detectors is not a sustainable solution, and not one we advocate). By involving students in the conversation about ethical use, we can hopefully instil a sense of responsibility that carries over to their professional lives, where AI will likely also be present.

\subsection{Privacy and Data Security}

A generic concern when deploying LLMs in educational settings is privacy. Many LLM services (like the free ChatGPT) operate on cloud servers – which means any student queries or content fed into them is potentially stored and could even be used to further train the model. This raises issues under student data protection regulations (such as FERPA in the U.S. or GDPR in Europe). Schools have to be cautious about what data (especially personally identifiable information or sensitive information) is shared with third-party AI services. For instance, if students are asked to input their essays into an online AI tool, is that essay now in the hands of the AI provider? Could it be regurgitated to another user later? These are valid worries. Some early incidents (outside education) saw AI chatbots reveal chunks of text from other users’ interactions due to system glitches. To address privacy, a few strategies are emerging:
\begin{itemize}
    \item Local or self-hosted models: Some establishments are exploring running LLMs on their own servers so that data never leaves their control. While currently the largest models are hard to self-host, smaller fine-tuned models could be used for certain tasks (e.g., an on-premises essay feedback model).
\item Data anonymization: Encouraging or mandating that students do not use real names or sensitive details when interacting with AI, or having an intermediary system strip personal info before sending a query to the LLM.
\item Vendor agreements: Schools can work with AI tool providers to ensure compliance with privacy laws, including assurances that data won’t be used beyond providing the service, won’t be kept longer than needed, etc. For example, some companies offer “education mode” AIs where conversations aren’t used to train public models and are kept confidential.
\item Transparency and consent: Parents and students should be informed when AI tools are being used that might involve their data, and consent should be sought where appropriate. In schoolchildren especially, parental consent might be needed for students to use certain AI services as per child online privacy laws.
\end{itemize}

Security is another facet – any system introduced (like a new AI app) increases the attack surface for hacks or data leaks. Schools will need to vet tools for security as well. For instance, if a poorly secured AI homework helper is breached, student submissions or feedback could be exposed. While not unique to AI, the more digital tools, the more security diligence required. Another aspect of privacy is intellectual privacy – the idea that students should feel free to ask questions or make mistakes without undue monitoring. If every interaction with an AI is recorded, one might worry about who can see that data. Will teachers or administrators be reviewing all AI chats? That could inhibit students from asking honest questions through fear of judgment. Some systems are designed with ephemeral interactions that are not permanently stored to alleviate this.

There is also the dimension of teacher privacy and academic freedom. If teachers use AI to draft materials, they might not want that to be scrutinized either (e.g., a teacher might privately use ChatGPT to get ideas for a lesson on a controversial topic such as  misogyny or young male attitudes to girls – if those queries are later exposed, could they be misinterpreted?). Ensuring that AI use remains a personal tool and not a surveillance mechanism is important for both students and teachers. These privacy and data considerations mean that educational institutions must be thoughtful and likely invest in infrastructure or agreements to use LLMs safely. Unlike casual consumer use, education involves minors and academic records, which are rightly held to higher standards of protection. This challenge is leading some schools to delay AI adoption until clearer policies are in place. However, it is an area where solutions are being actively worked on, and frameworks (like Wang et al.’s mention of “privacy and security” as a key challenge\cite{Wang2024Survey}) emphasize building in these protections from the start. Ensuring trust in AI tools will heavily depend on guaranteeing they don’t compromise user privacy.

\subsection{Transparency and Explainability}

As LLMs take on roles in education, their opacity becomes a concern. These models are often “black boxes” – they give answers without explaining their reasoning or knowledge sources. In an educational context, this lack of transparency can be problematic. For one, if an AI tutor gives an explanation, a student might ask “Why should I trust that?” or “How did you get that answer?” The most recent  LLMs are addressing this by having a reasoning or thinking mode, in which the major steps they follow in problem breakdown and processing are exposed to the user.  Teachers and parents might want to know on what basis the AI is giving certain advice to a student; if an AI recommends a student focus on a certain topic (maybe because it inferred a weakness), what evidence is that based on?

Explainability is also linked to trust. Research suggests that when AI systems can provide explanations or rationales for their outputs, users trust them more and find them more useful, especially when the explanation helps diagnose if the AI might be wrong. In education, an explainable AI could, for example, highlight which parts of a student’s essay led it to give a certain feedback comment, or it could say “I solved the equation by applying the quadratic formula” rather than just giving the answer. Some recent work is attempting to incorporate self-explanation features in educational LLM applications. For example, an AI maths tutor might be configured to always produce a step-by-step solution (which the student can then compare to their own steps). This is a form of transparency that doubles as pedagogy. Another angle is using techniques from interpretable machine learning: simplifying the model’s decision to something like a set of rules or extracting key patterns. For instance, if an AI classifier is used to flag student essays that might need teacher intervention, it could highlight the phrases or features that led to the flag (like “off-topic” or “inappropriate language”). A related concern is accountability: if an AI gives bad advice and a student is misled, who is accountable? Transparency can help trace the issue – was it a misunderstanding in the prompt, or a flaw in the model, etc.? Without transparency, it is hard to even diagnose problems. One illustrative scenario: Suppose an AI tutor consistently gives less detailed answers to questions about women scientists than men (reflecting a bias in the training data). If we don’t have tools to analyze its outputs systematically, such patterns might go unnoticed, quietly disadvantaging some content, but if the system logged and could aggregate how it responds, teachers/developers might catch and correct that. So, building some level of monitoring and explainability into educational AI deployments is critical. Explainability is also important for getting teacher buy-in, as many teachers are rightfully wary of using AI if they feel they can’t understand or control it. If the AI can provide explanations or allow the teacher to inspect its knowledge base (e.g., show what reference material it’s drawing from), the teacher may feel more confident integrating it. Kasneci et al. (2023)\cite{Kasneci2023} argue that teachers and learners need to develop “competencies and literacies to understand the limitations and unexpected brittleness” of such systems. Part of that understanding comes from the system being as transparent as possible about what it can or can’t do. In current practice, a simple approach to transparency is for AI systems to clearly state their limitations to users. For example, some educational chatbots begin interactions with a note like “I am an AI tutor. I might sometimes be wrong or incomplete. Always double-check my answers and ask for clarification if needed.” While this doesn’t explain individual answers, it sets the frame that the AI is a fallible assistant, not an oracle. Another practice is logging the AI-student interactions for review: teachers or the system maintain a transcript that can be reviewed if issues arise. This transparency of interaction can help in mentoring the AI’s behaviour too – essentially providing data for improvement. However, we must balance this with the earlier privacy concern; such logs should be kept secure and only used appropriately.

A more technical notion of transparency is algorithmic transparency: informing stakeholders about how the LLM was trained, what data might have influenced it, what version is being used, etc. If a school uses an AI model, they should ideally know if it was trained on general internet data up to 2021, or if it has some fine-tuning on educational content, etc. This context helps interpret its outputs and limitations. In regulated sectors like healthcare, there is often requirement to use algorithms whose functioning is somewhat interpretable; education might head that way if AI becomes deeply embedded. Future research is likely to focus on “opening the black box” of LLMs in education, possibly with hybrid systems that combine neural nets with more interpretable logic or knowledge components. Until then, educators and AI developers should strive to make the AI’s operation as visible as possible to its users, and cultivate an environment where questions like “why did you say that?” are encouraged – whether asked to the AI or about the AI.

\subsection{Teacher Adoption and Training}

Implementing LLM-based tools in education also faces pragmatic challenges of teacher adoption and training. No matter how good an AI tutor or content generator is, if teachers do not understand it or feel comfortable with it, it will not be used effectively (or at all). Many teachers today have little to no formal training in AI and the sudden rise of LLMs has in some cases left educators feeling overwhelmed or concerned about their place in the classroom. Some may see AI as a threat to their expertise or worry that relying on it could diminish their role, and overcoming these perceptions requires proper professional development and support. Initial experiences suggest that when teachers are given hands-on training and time to experiment with LLM tools in a low-stakes setting, they often become more open to using them. For example, a school district might run workshops where teachers try out ChatGPT for their own tasks (like generating a lesson plan) and discuss the results. By seeing the AI as a personal aid first, teachers can build intuition and reduce fear. Another aspect of training is pedagogical integration: teachers need models of how to incorporate LLMs into their teaching practice. This could include example lesson plans that use AI, guidelines for facilitating class with AI, and strategies for handling common issues (like a student getting a wrong answer from AI – how to turn that into a teachable moment). Early adopter teachers are already sharing such practices in communities of practice (for instance, via blogs or education conferences). Teacher education programs (both pre-service and in-service) are beginning to add modules on AI in education. A survey by Zhang \& Aslan \cite{Zhang2021} found that while most teacher training curricula did not yet cover AI, there was strong interest among teachers for such training, indicating a gap that needs filling. One proposal is to treat AI competency as a new component of digital literacy for educators – alongside knowing how to use productivity software or online learning platforms, teachers will need to know how to leverage AI tools and understand their implications. Additionally, support from school leadership and policies will influence teacher adoption. If administrators encourage AI experimentation and provide resources (like subscription to a better AI service, or time for teachers to develop AI-enhanced materials), teachers are more likely to engage. Conversely, if policies are unclear or only emphasize the negatives (e.g., “don’t let students use AI”), teachers might steer clear of integrating it positively. A supportive culture might involve, for example, an AI learning community on campus where teachers regularly meet to share experiences, or pairing tech-savvy teachers with those less comfortable for peer mentoring. Also worth noting is that not all educational settings have the same levels of support – for instance, many higher education instructors have autonomy but not necessarily institutional support or training, so they rely on self-learning or informal networks. Meanwhile, schoolchildren's teachers might have more developmental opportunities but also more regulations to follow. Bridging this, some major education conferences (like ISTE or Learning Technologies) have started offering tracks on AI in education specifically aimed at practitioners, which helps disseminate knowledge. Finally, time is a non-trivial challenge: teachers are extremely busy, and learning a new technology and redesigning lessons around it takes time and effort. Without clear payoff or compensation (or reduction in other duties), expecting teachers to magically integrate AI on top of everything else could lead to burnout or resentment. Showing that AI can reduce some burdens (like grading load or planning time) is important to win buy-in. And indeed, surveys (e.g., Labadze et al.\cite{Labadze2023ChatbotsSLR} highlight “time-saving” as a key benefit of AI chatbots for educators. Thus, framing training in terms of how AI can make teachers’ lives easier, rather than just another thing they have to implement, is prudent.

These challenges underscore that introducing LLMs into education is not a plug-and-play solution; it involves complex trade-offs and design choices.  Wang et al. \citeyear{Wang2024Survey} categorize many of these risks including fairness, safety, transparency, privacy, and over-dependence. Shahzad et al.\citeyear{Shahzad2025Comprehensive} similarly emphasize balancing ethical and pedagogical considerations alongside personalization benefits. What is encouraging is that none of these challenges appear insurmountable – rather, they require thoughtful policies, improved technology, stakeholder education, and continuous evaluation. The field is actively working on solutions, from technical fixes (like bias reduction algorithms) to pedagogical frameworks (like clear usage policies and ethics training). 

 It is clear that investing in teacher capacity-building is integral to successful LLM adoption in education. This is a sociotechnical issue: it is not just about building the tool, but building the human infrastructure around it. Teachers empowered with knowledge and skills about AI are more likely to harness it in innovative ways that truly enhance learning. Without that, even the best AI could languish unused or be used suboptimally. The effectiveness of AI in classrooms will depend less on the AI itself and more on the teacher who wields it.

\section{Reviewing the Pace of Change}

The intersection of large language models and education has evolved at a breakneck pace over the past few years. What was a speculative idea not long ago – AI systems that could hold meaningful educational conversations – is now a reality being piloted in classrooms worldwide.

\subsection{Recent Milestones}

In the pre-LLM era, the domain of AI in education was dominated by intelligent tutoring systems (ITS), adaptive learning software, and simple chatbots. These were often domain-specific (e.g., a tutoring system for algebra that followed a scripted model of common student errors) and had limited natural language ability. Earlier surveys, e.g. Holmes et al.,\citeyear{Holmes2019} chronicled steady progress in ITS effectiveness, but also the difficulty of scaling to open domains. The emergence of deep learning-based language models like BERT and GPT-2 (2019) hinted at new possibilities for more generalist AI tutors, but it wasn’t until GPT-3 (2020) that the capabilities truly leapt forward to handle a wide array of educational queries. Even then, early research on GPT-3 in education was tentative – small-scale experiments or thought pieces. The watershed moment was the release of ChatGPT in late 2022, which for the first time put a powerful LLM directly into the hands of millions, including educators and students. Within months, anecdotal reports of its use in classrooms and homework circulated widely, prompting academic investigations. By early 2023, a flurry of position papers and editorials appeared, aiming to grapple with the sudden impact: e.g., Latif et al., 2023 \cite{latif2023artificial} called ChatGPT a wake-up call for educators, while some, like Weismann 2023, provocatively titled articles as “ChatGPT is a plague upon education”\cite{Weismann} reflecting alarm. 

Recognizing the need for clear-eyed analysis, Kasneci et al. (2023)\cite{Kasneci2023} published their commentary “ChatGPT for Good?” – it can be seen as one of the first comprehensive looks at both opportunities and challenges, setting an agenda for required competencies such as critical thinking and fact-checking in the face of generative AI. Soon after, more systematic scholarly efforts began: Bhullar et al. (2024)\cite{Bhullar2024ChatGPTHE} conducted a bibliometric analysis of the burgeoning literature on ChatGPT in higher education, identifying major themes ( academic integrity and student engagement). Ansari et al. (2023)\cite{Ansari2023ChatGPTScoping} completed a scoping review of global studies, revealing patterns in how ChatGPT was being used by teachers and students, with a key insight that most uses aimed at reducing workload or getting quick assistance. By late 2023, initial empirical studies and small-scale experiments were reported: e.g., studies measuring student learning outcomes with AI assistance in specific tasks, or surveys of student/teacher attitudes after trying AI in a course. The meta-analysis by Deng et al. (2024)\cite{Deng2024Meta} synthesized results from 69 experimental studies – notable because it indicates that within about a year of ChatGPT’s rise, dozens of controlled studies had already been done, something quite rapid for educational research. They found generally positive effects on performance and motivation, but also pointed out methodological limitations in those early studies (small samples, short-term). Meanwhile, the AI and HCI communities produced design-oriented research: for instance, Wang et al. (2024)\cite{Wang2024Survey} and Xu et al. (2024)\cite{Xu2024Survey} compiled surveys looking at technological advancements and proposing classifications of LLM educational applications (like the toolkit of chatbots, quiz generators, etc., previously discussed). These works also emphasize future research needs, such as specialized educational datasets and benchmarks to evaluate LLMs’ pedagogical skills. On the practitioner side, by 2024 we see organizations like ISTE, EDUCAUSE, etc., releasing guidance documents and whitepapers on AI in education. Governments too have shown interest: the U.S. Department of Education published a report “AI and the Future of Teaching and Learning” \cite{Cardona2023} outlining opportunities and cautioning on risks. 

This interplay of top-down guidance and bottom-up experimentation is shaping a more mature view of LLMs in education. An interesting development is the inclusion of sociotechnical perspectives early in the conversation. Unlike some past ed-tech waves that focused heavily on efficacy before ethics, the AI wave has had ethics, equity, and teacher roles as central topics from the start – likely because of the broad societal attention on AI. For instance, prior survey papers like Labadze et al. (2023)\cite{Labadze2023ChatbotsSLR} explicitly review not just benefits but also limitations, concerns, and prospects, giving weight to issues like reliability and teacher acceptance. Shahzad et al. (2025)\cite{Shahzad2025Comprehensive} proposed a theoretical framework with ethical and pedagogical balance as one pillar, indicating that a holistic approach is being pursued. In terms of adoption in practice, the field has moved from initial pilots to more structured implementations. For example, Khan Academy’s pilot of Khanmigo in 2023 with thousands of students provided a lot of real-world feedback (some of which has been shared in blogs and interviews). Several universities ran controlled experiments, like MIT’s trial of an AI coding tutor in an introductory programming class (reporting that while it helped with syntax issues, students still needed conceptual guidance from humans)\cite{ma2024integratingaitutorsprogramming}. By 2024, some institutions began drafting AI use policies for students (e.g., allowing AI with citation). My own institution has a variety of policies, with one programming project module allowing students to choose from one of four positions: no use of AI; partial use to help problem-solve; full use including coding help, or their own definition of ethical use for their own group.  What is less clear is how this impacts effective assessment, though it has the benefit of being clear to the students what could be acceptable or not, and giving them autonomy to make the choice.  Competitions and hackathons emerged to build educational apps on top of GPT APIs, generating a slew of tools (though many are short-lived startups). Notably, the field has also begun to connect with classic learning theories and practices: for instance, linking LLM use with Vygotsky’s concept of the Zone of Proximal Development (ZPD)\cite{vygotsky_mind_1978}, where the AI can act as the “more knowledgeable other” to scaffold learning just beyond the student’s current ability, or exploring how AI fits into social constructivist models (as a participant in knowledge construction). This theoretical grounding helps move beyond hype to deeply considering how learning happens with AI.

\subsection{Summarizing the evolution}
\begin{itemize}

\item Phase 1 (2022-early 2023): Enthusiasm and alarm, largely opinion pieces, early adopters testing the waters.  Lots of unawareness.
\item Phase 2 (mid/late 2023): Rapid emergence of empirical studies, literature syntheses, and the development of guidelines; initial classroom integrations and case studies. Some apathy.
\item Phase 3 (2024): More rigorous research (longitudinal studies, larger samples, cross-cultural studies), development of specialized models and tools, integration into curricula in more structured ways, and refinement of policies at institutional and governmental levels. Rising awareness.
\item Phase 4 (2025): Introduction of reasoning LLMs with increased capabilities and apparently increased transparency.  Improvements in reducing hallucinations and errors. but by no means completely reliable for expert information. Broader discussion.
\end{itemize}
The field is still very much in flux, but it is maturing quickly, drawing on lessons from these early experiences.

\section{The Impending Revolution: Implications for the Design of  Technologies} 
In this section we develop arguments about some of the larger changes that are coming: one of the biggest impacts on education is likely to come from a shift in learner expectations owing to an increased exposure to LLMs, whether in the classroom or at home.  These shifts are hardly apparent at present, owing to the novel nature of the technology and its limited use so far, but we expect these factors to become significant in the future.  We postulate that these will have a significant impact on the design of future technologies and systems, and in the field of ed-tech in particular.

\subsection{Changing the default interaction style}
The scroll-point-click (windows; icons; mouse; pointer - WIMP on laptops; scroll/touch on smartphones) interface style is the predominant interaction paradigm for much of current technology, prevalent across laptops, desktop and mobile interaction.  It occurs in most domains, and many users have never really interacted with anything else such as a command line.  The widespread usage of LLMs is likely to create a significant driver for change that could see these sorts of systems becoming much less common.

Consider a current scenario where a learner is finding out about the history of anaesthetics.  They read material on the screen, scrolling to access longer text below the page fold.  They click on links in the text to go to further material, or select icons that take them to alternative details.  If what they want is not there, then after looking around for it, they have to find a search bar and type in, for example "Lister and carbolic spray" and obtain a page of search results on that topic, nowadays lead by an AI summarisation of key aspects.  They then scroll and click to continue their exploration.  Thus, in a WIMP system, users have to identify what they need on the screen; if not there, they need to scroll or change screens to access it, or find an entry point to be able to enter text to redefine what is on the screen, typically via search.

With an LLM, the interaction is completely different: it is conversational. This can be via speech, or via text.  A shift towards speech-based interaction has been a long time coming, and devices such as Alexa and Google Home have accelerated its takeup, but until now it has been the preserve of small command sets or specific focussed scenarios such as in operating theatres.   And whilst speech-based input is a useful advantage of LLMs this is not the major pivot.

The key change is that the screen becomes non-referential.  The location of items on the screen is unimportant, and if we want to explore something on there we refer to it by content, not location --- "\textit{give me more on the role of carbolic acid in killing germs"}.  If what we want is not available, we simply use the same interaction point to ask for more information: there is no need to move in a searching mode.  The second change is that the interaction becomes more organic and free-flowing --- learners are aware they can ask whatever they wish and then refine their enquiry based on what they see: the cognitive processing required to express their next steps is much lower since what they they simply say or type what they need in their native language, rather than having to interpose it with moving a pointer and clicking in various locations\cite{abowd1991users}.  Thus the interaction becomes a much more natural discourse between computer and user, with correspondingly lower costs of doing the wrong thing. 

Combined, the lack of need to involve pointers and clicks, and the conversational refinement of exploration, these factors provide a very natural, low-barrier to entry and accessible interaction style, and it seems reasonable to assume that users will shift towards preferring this simple, less cognitively demanding approach. The evidence for this is in the rapid take=up of LLMs in education generally, and in user experiences of them in other scenarios.  It is currently the case that few people have fully experienced ongoing interactions with LLMs, but their spread suggests that within short order, this situation will be reversed.  This has significant implications for the design of much ed-tech.  Our claim is that uses will move from the WIMP interface and demand conversational interfaces, meaning many of the existing approaches may have to be modified to satisfy user expectations and needs.  Whilst it can be argued that different styles of material are helpful for a variety of different purposes, the fluidity and breadth of the conversational interaction style will become a preferred mode of learning.

\subsection{Context awareness}
LLM systems, at least the most up to date ones, have a context awareness that spreads a thread of conversation.  This can exist for a short interaction only, or over a longer period.  This allows the system to respond appropriately when asked about certain things, encouraging deictic references and more natural conversations.  Most existing interactive systems only have a very limited notion of context, and so often require repeated instructions to retain the focus on the particular topic of interest.  This becomes frustration: if you are searching in Google for how to achieve some video effect in Final Cut Pro, you have to keep adding 'Final Cut Pro' to your search terms as you home in on the answer, but with an LLM it becomes part of the relevant context and makes locating information much simpler.   Having context extend over much longer periods is also something relatively unfamiliar to users in existing systems, where new sessions tend to start from scratch again.  This does not reflect how learning works or how the education system runs: we return again and again to topics, scaffolding learning on previously understood concepts.  To effectively start afresh on a new session is counter-intuitive. Thus, because an LLMs context extends to understanding what the learner has covered and knows about, it can adjust the responses accordingly, making it a much more useful educational partner. 

\subsection{Personalisation}
The awareness of current context also extends to such an extent that it can be considered personalisation.  The LLM can be given prior information on the learner and how to work with them, such as "treat the learner as an intelligent but uninformed 15 year old, who prefers descriptive examples to logical formalisms, and encourage them to explore ideas by being supportive and not critical of any misunderstandings".  This allows for a bespoke form of personalisation based on pedagogical considerations for an individual.  The history of conversation and questions answered correctly and those areas for improvement all become part of the inherent model of the user, providing a unique tailored experience.  We are in an era where Bloom's notion of individual support to allow excellence is becoming a reality.   Future work must reconcile AI’s ability to personalize pathways with the need to cover agreed-upon learning goals. The “Mixture-of-Experts” framework mentioned in Razafinirina et al. (2024)\cite{Razafinirina2024Personalized} is one idea: different specialized mini-AIs for different subjects under a central orchestrator,  ensuring that while personalization happens, it is still anchored to expert knowledge in each domain.

We also posit that when this personalisation becomes the expectation and norm in LLM learning, it is unwise to assume users will readily give this up and return to more ignorant, less aware systems for other forms of interaction, and so there is likely to be a need for designers to shift towards understanding and accommodating individual differences and needs.  

\subsection{Generality}
Learner experiences of LLMs show that they enjoy the exploratory nature of their interactions, and the LLMs  are not limited to answering questions on just the topic in question (unless specifically directed to be highly focussed in the prompts).  This allows for serendipitous exploration, diving down rabbit holes of interest, as well as displacement activities and non-work.  Their expectations will be shaped by this almost boundary-less nature, which, when coupled with the conversational style, is likely to become a default.  Systems that are specific to one thing, inflexible and unable to respond to things slightly off-topic, will be viewed with frustration and see a lack of engagement and takeup.  Thus, there is an imperative for designers to consider when having walled explorations in focussed spaces is sensible and how to accommodate the freer style of many other forms of interaction.

We hypothesise that ed-tech developers will increasing provide modules for specific purposes or pedagogical approaches that fit into an ecosystem of components that include LLMs.  This helps address the design push towards generality: not everyone wants to include an LLM into their own systems, and trying to combine knowledge from multiple LLMs is still technically challenging.  This represents a transition from the app-centric perspective where all interaction, data and learner information is combined into one islanded system. 

\subsection{Trust through explainability}
We have already discussed explainability but more from the perspective of the educator.  As LLMs provide more information about their reasoning, and can be interrogated conversationally about why they provide answers, users are able to build up a picture as to why they have given the answers that they have (whether correct, or incorrect).  Because of this they can build a picture of what they can trust and what to be wary of from the system.  There is only need for trust if there is risk: there is clear risk for learners in not achieving their goals, understanding things they want to or need to for their exams.  If they have got used to trusting systems because those systems can explain themselves, they are likely to be less tolerant of other systems that just present information and are opaque as to how they are providing answers and are unable to be self-reflective. Again, this will drive a groundswell of opinion away from working with the more conventional systems currently in place: useful and effective as they may be, if the users feel that information is provided without knowing why or where it comes from or being able to query and check it, they will necessarily trust it less and so be less inclined to engage with it.

\section{Future Research Opportunities}

Looking ahead, several trends and needs for future research and development are apparent from the literature.

\emph{Multimodal and More Capable Models}
Thus far, we have mostly discussed text-based LLMs. However, the more recent LLMs can manage to process speech effectively, and respond in a variety of voices too.  This can be very engaging ---- I created a LLM that acts as a dream interpreter, giving Freudian or Jungian perspectives on dreams that people tell it, and it is interesting to see the engagement that chatting to this system generates in people.  Generative models that can output images, audio, or video in addition to text are available, though have distinct limitations (at the time of writing, generative AI cannot produce an image of a left-handed person, for example).  Their power is impressive. however: I recently asked it to produce a video showing Shakespeare writing Macbeth,  and within minutes had a CGI-level video depicting Shakespeare in his study, narrating how he created Macbeth complete with cutaway shots of Macbeth, the witches and Lady Macbeth playing cameo roles,  all created for free (environmental costs aside) and within minutes, with no artistic talent required.  In education, this could unlock AI tutors that can see a student’s work (like a photo of a maths solution they wrote) and provide feedback, or language learning assistants that not only chat but also speak and listen (improving pronunciation). A recent survey by Wu et al. (2023)\cite{wu2023multimodal} specifically calls attention to multimodal LLMs in education, in which images can be used alongside music, speech and text, suggesting future systems where, for example, an AI tutor could explain diagrams and annotate them interactively with the student, or a virtual lab assistant can both talk and conduct a realtime simulation experiment, accompanied by video analysis of the results.

\emph{Frameworks and Theories for Integration}
We expect to see more comprehensive pedagogical frameworks emerge that integrate LLMs. Shahzad et al. (2025)\cite{Shahzad2025Comprehensive} offered one with three pillars (personalization, ethical balance, adaptability). Others may propose frameworks clarifying which levels of cognitive skill AI should be used for and which should be left to humans\cite{Huang2018}, or models of the teacher-AI-student triad that optimize learning.

\emph{Longitudinal and Transfer Studies}
A lot of current studies are short-term. Future research should look at long-term effects: e.g., follow a cohort of students who use AI heavily throughout high school – how do their skills, habits, and outcomes compare to those who did not, when they reach college or jobs? Does continuous use lead to dependency or does it free up time to learn more deeply? Also, studying whether skills learned with the help of AI transfer to situations without AI (do students become better self-regulated learners, or conversely, do they struggle without AI prompts?). These will inform guidelines on how frequent AI usage should be, and when to fade support.  Long-term studies will also be needed to understand the shift in user expectations regarding design approaches and their attitudes to the use of other systems.

\emph{Evaluation Metrics and Benchmarks}
The community will likely develop standard benchmarks to evaluate educational LLMs. For example, tasks like “explain a grade-level science concept accurately and pedagogically helpfully” or “diagnose a common student error from their response and give a helpful hint” could become benchmark tasks that different models are tested on\cite{Wang2024Survey}. By benchmarking, we can track progress and identify which models or methods work best for educational purposes, as a model great at general chat might not be the best teacher out-of-the-box.

\emph{Ethical and Policy Research}
As implementation grows, we will see real-world data on things like academic integrity violations, bias incidents, etc., which will allow researchers to study these empirically and refine policies. For instance, do strict AI bans actually curb cheating or just drive it underground? Are there effective honour code pledges or monitoring systems that maintain trust? How do students from different backgrounds perceive learning with AI - who finds it most helpful vs who finds it frustrating? Sociological and psychological research will deepen our understanding of the human factors at play.  We will also need to understand the role of user-derived data in training, since many models take the information provided by learners to enhance the training of their models.

\emph{AI Literacy and Curriculum Changes}
On the curricular side, one likely direction is formally incorporating AI literacy into education. Already some schools are adding units on AI – how it works, its impacts, how to use it responsibly, and this will likely become mainstream, just as internet literacy became part of education. Having students build simple prompts or small chatbot programs could become a way to demystify AI. Organizations like the ISTE have started drafting AI literacy standards for students. Coupled with that, core curricula might shift to emphasize more what humans do best – creativity, critical thinking, collaboration – leveraging AI to handle rote tasks. In effect, the presence of AI could push education towards the higher-order skills which have always been a goal.

\emph{Collaboration between AI developers and Educators}
Another future direction is more interdisciplinary collaboration. Up to now, many education researchers have had to react to AI tools developed by industry for general use. Going forward, we might see more custom models or tools co-developed with educator input from the ground up; for example, fine-tuning LLMs on transcripts of human tutoring sessions to teach it tutoring style, or ed-tech companies partnering with universities to trial new forms of AI-driven pedagogy (some such partnerships are already in early stages\cite{newman_edtech_2023}). This collaboration will ensure that future AI is designed for education, not just repurposed for it.

\emph{Addressing Limitations of Current Models}
On the AI research side, we expect continued work to mitigate issues like hallucination and lack of reasoning transparency. Techniques like integrating symbolic reasoning (so the AI can explicitly solve equations or logical problems stepwise) could greatly benefit its reliability in subjects like maths, though recent models have come on dramatically in this area. There is also work on making models smaller and more efficient (so schools could run them locally, addressing cost and privacy). Edge computing and on-device models could allow offline or low-latency use in classrooms without constant internet – one of Wang et al. (2024)’s future directions was efficiency and edge deployment\cite{Wang2024Survey}. If realized, that could dramatically increase accessibility globally.

\emph{Interdisciplinary Learning and AI}
As AI can connect different fields of knowledge easily, future educational models might break down the subject silos. For instance, an AI could help a project that involves history, maths, and art simultaneously, encouraging a more interdisciplinary learning approach. Research could examine how AI might foster connections between disciplines for students, something human teachers strive for but can be challenging to coordinate.  Ultimately, the long-term future might see the concept of an AI-augmented classroom become standard: where every student has an AI assistant tuned to their needs, the teacher orchestrates learning with real-time data from AI analytics, and the curriculum is dynamically adaptable. It is a vision where AI is ubiquitous yet  fades into the background as a supportive infrastructure, letting the human elements of curiosity, creativity, and interpersonal interaction take centre stage in learning. 

It is clear that the field of LLMs in education is in an exciting formative period. Early evidence shows significant promise: improved personalized learning, increased engagement, and new instructional possibilities, and is likely to drive new demands in interactional styles from learners and teachers alike.  At the same time, it has brought to the forefront important conversations about the purpose of education in an AI-rich world (indeed, the purposes of education itself) –-- prompting re-examination of what we teach and how we assess when information and even cognitive processes can be automated. As Labadze et al. \citeyear{Labadze2023ChatbotsSLR} and others summarize, the consensus is that LLMs are not likely to replace educators but can radically empower those who learn to work with them. Education has always evolved with technology (from books to calculators to computers), and LLMs represent another leap –-- one that could bring us closer to the age-old dream of truly personalized, accessible, and engaging education for all, if we navigate the journey with empathy and care. 

\emph{Acknowledgments}
The concepts relating to the design challenges, learner expectations  and interaction issues that these developments will lead to are all the work of the author.  This was grounded in a literature review that drew on a range of recent studies and surveys to provide a comprehensive synthesis.  It was produced by the author working collaboratively with generative AI to explore specific areas through prompt engineering and conversational exploration, evaluating and checking information, which was then rewritten, edited, restructured and refined multiple times.  
\bibliographystyle{acm}
\bibliography{references}

\end{document}